\documentclass[aps,prb,preprint,onecolumn,superscriptaddress]{revtex4-2}

\usepackage{amssymb,amsmath,amsfonts}
\usepackage{graphicx}
\usepackage{float}
\usepackage{natbib}
\usepackage{color}
\usepackage{lmodern} % improves the default font
\usepackage[colorlinks=true,bookmarks=false,citecolor=blue,urlcolor=blue]{hyperref}
\usepackage{tikz}
\usepackage{wrapfig}

\graphicspath{ {./Figures/} }

\newcommand{\br}{\mathbf{r}}
\newcommand{\bk}{\mathbf{k} }

\newcommand{\bE}{\mathbf{E} }

\newcommand{\bH}{\mathbf{H} }

\newcommand{\bj}{\mathbf{j} }
\newcommand{\bP}{\mathbf{P}  }
\renewcommand{\Re}{\frak{R}\mathrm{e} }
\renewcommand{\Im}{\frak{I}\mathrm{m} }

\newcommand{\dyad}[1] {\overset\leftrightarrow{\mathbf{#1}}}
\newcommand{\dyadA}{\overset\leftrightarrow{ \alpha }}
\newcommand{\dyadtildeA}{\overset\leftrightarrow{ \tilde{\alpha} }}

\newcommand{\eps}{\varepsilon}
\newcommand{\w}{\omega}

\newcommand{\bkp}{\bk_{\parallel} }

\renewcommand{\tilde}{\widetilde}

\begin{document}

\newcommand{\qingdao}{
Qingdao Innovation and Development Center of Harbin Engineering University, Qingdao 266000, Shandong, China
}

\newcommand{\mipt}{
Center for Photonics and 2D Materials, Moscow Institute of Physics and Technology, Dolgoprudny 141700, Russia
}

\newcommand{\itmo}{ITMO University, St. Petersburg 197101, Russia}

\title{Emergence of collective spectral features in finite arrays
of dielectric rods}

\author{Ilya Karavaev}
\affiliation{\qingdao}
\affiliation{\itmo}

\author{Ravshanjon Nazarov}
\affiliation{\itmo}

\author{Yicheng Li}
\affiliation{\itmo}

\author{Andrey A. Bogdanov}
\affiliation{\qingdao}
\affiliation{\itmo}

\author{Denis G. Baranov}
\email[]{baranov.mipt@gmail.com}
\affiliation{\qingdao}
\affiliation{\mipt}

\begin{abstract}
Periodic optical structures, such as diffraction grating and numerous photonic crystals, are one of the staples of modern nanophotonics for the manipulation of electromagnetic radiation. The array of subwavelength dielectric rods is one of the simplest platforms, which, despite its simplicity exhibits extraordinary wave phenomena, such as diffraction anomalies and narrow reflective resonances. Despite the well-documented properties of infinite periodic systems, the behavior of these diffractive effects in systems incorporating a finite number of elements is studied to a far lesser extent. Here we study theoretically and numerically the evolution of collective spectral features in finite arrays of dielectric rods. We develop an analytical model of light scattering by a finite array of circular rods based on the coupled dipoles approximation and analyze the spectral features of finite arrays within the developed model. Finally, we validate the results of the analytical model using full-wave numerical simulations.
\end{abstract}

\maketitle
\newpage

\section{Introduction}

Periodic optical structures, such as diffraction grating and various photonic crystals, offer a wide range of optical phenomena, such as optical band gaps, hyperbolic dispersion, negative refraction, and others \cite{DeAbajo2007, Ebbesen1998, Martin-Moreno2001}. 
This diversity of possible optical behaviors, studied since the pioneering works of Wood and Rayleigh \cite{rayleigh1907iii, wood1902xlii}, makes them one of the staples of modern nanophotonics for advanced light manipulation \cite{Auguie2008, Manjavacas2019}.
Ultimately, the resonances of such arrays enable optical bound states in the continuum (BIC) \cite{Marinica2008,zhen2014topological, koshelev2023bound}.
In real structures BICs can be detected as resonant states with sufficiently high quality factors \cite{joseph2021bound}. Thus, the theoretical and experimental investigations of the BICs have opened up novel avenues  in the study of topologically protected states \cite{kang2022merging}, creation of lasers \cite{hwang2022nanophotonic} and highly sensitive dielectric sensors \cite{kang2023applications}. 
%Such states can be  characterized as a polarization vortex  at momentum space, the winding number of which determines their topological charge \cite{zhen2014topological}. The presence of topological guarantees the robustness of BICs against perturbations, which do not change the in-plane symmetry of the structure\cite{koshelev2023bound}. In real structures BIC can be detected as a resonant state with a sufficiently high quality factor\cite{joseph2021bound}. Thus, the theoretical and experimental investigations of the BICs have opened up novel avenues  in the study of topologically protected states \cite{kang2022merging}, creation of lasers \cite{hwang2022nanophotonic} and highly sensitive dielectric sensors\cite{kang2023applications}.  

Particularly, periodic arrays of dielectric rods present an ultimately simple optical system that, nevertheless, is capable of exhibiting some extraordinary wave phenomena \cite{Borisov2005}. For instance, in article \cite{maslova2021bound}, authors investigated the appearance of BIC within a low-contrast bilayer resonator consisting of an infinite array of dielectric rods and theoretically discussed the stability of these states against the structural disorder. Previous theoretical and experimental studies have shown remarkable spectral features in arrays of thin dielectric rods, such as ultra-narrow resonances and total transmission \cite{gomez2006extraordinary, Laroche2006a, Laroche2007, Du2013}.
These effects have been attributed to the far-field inter-particle interactions in the system in the vicinity of so called diffraction, or Rayleigh, anomalies \cite{DeAbajo2007}. 
Interestingly, total transmission in an array of thin rods occurring exactly at the Rayleigh anomaly remains even in the presence of material losses in the dielectric rods.
Although these and related effects have been widely studied and verified experimentally \cite{Ghenuche2012} in periodic structures, one can wonder how these diffractive effects emerge in finite systems incorporating only a finite number of scattering elements.

In this Letter, we theoretically investigate evolution of collective spectral features in finite arrays of dielectric rods. We develop an analytical model of light scattering by a finite array of circular rods based on the coupled dipoles approximation (CDA). 
Using this model, we examine the effect of the number of the elements of a finite array, and the material absorption on the extinction spectra of the array.
Finally, we verify the results of the CDA-based analytical model by performing numerical simulations of he finite system.

\section{Results}
\begin{figure}[t!]
\includegraphics[width=.8\textwidth]{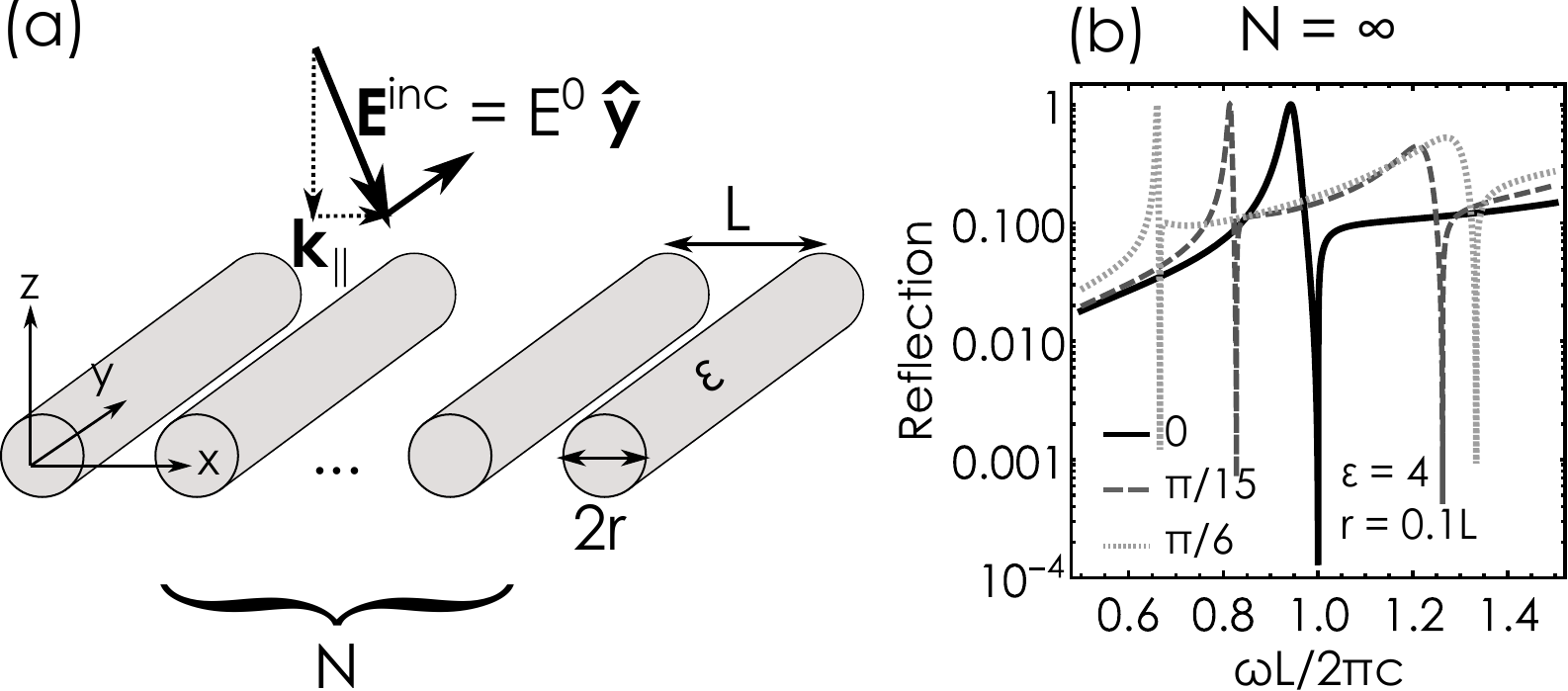}
\caption{(a) Geometry of the system: an array of $N$ parallel infinitely long circular cylinders with permittivity $\eps$, radius $r$, and distance $L$ between the centers of adjacent rods.
(b) Reflection spectra of an infinite periodic array of dielectric cylinders ($\eps = 4$, $r = 0.1 L$) under TE-polarized illumination at angles $\theta = 0, \pi/15, \pi/6$.}
\label{fig1}
\end{figure}

Figure \ref{fig1}(a) illustrates the system under study. We consider a system of $N$ parallel infinitely long circular dielectric cylinders with permittivity $\eps$ and radius $r$ and distance $L$ between the centers of two neighboring rods.

To describe the general spectral features of the system, we quickly overview the response of the ideal periodic array, $N = \infty$.
Figure \ref{fig1}(b) presents intensity reflection spectra of an infinite array for a series of incidence angles $\theta$. 
The spectra clearly display two key features of the dielectric rods array. First, the array becomes transparent at the Rayleigh anomaly   \cite{gomez2006extraordinary} defined by
\begin{equation}
    \left| \bkp  + \frac{2\pi m}{L} \right|= \frac\w {c},\quad m \in Z,
    \label{eq.ref1}
\end{equation}
and corresponding to the diffraction channel opening.
Another interesting feature is observed at frequencies slightly below the first lattice singularity, where the array becomes fully reflective, $|r|^2 = 1$. 
These reflection peaks are a manifestation of the underlying quasi-normal modes of the array \cite{Borisov2005}.
%Figure Sx of Supporting information shows the same spectra but as a function of both frequency and the in-plane momentum $\bkp $ of the incident field. 
Non-zero in-plane momentum $\bkp  \ne 0$ lifts the degeneracy between $+n$th and $-n$th diffraction orders, whose onsets now occur at different frequencies. At each Rayleigh singularity the array becomes fully transparent.

\subsection{Effect of finite size}
Now we move to to a system containing a finite number of the rods.
Consider array of $N$ scatterers can be described by point sources located at $\br _1, \br _2, ... , \br _N$. Here we utilize original Foldy-Lax formulation of scattering problem in coupled dipole approximation:% assuming following behavior for total field
\begin{equation}
    \bE (\br ) = \bE ^{\mathrm{inc}}(\br ) + 4\pi k_0^2 \sum\limits_{j = 1}^{N} \dyad{G}(\br , \br _j)\bP (\br _j),
    \label{eq.ref2}
\end{equation}
where $\bP _j \equiv \bP (\br _j) = \dyadA_j \bE (\br _j)$ is the dipole moment of the $j-$th scatterer defined by self-consistent exciting field, and $\dyad{G}$ is the free space dyadic Green's function of the 2D Helmholtz equation:
%satisfying Helmholtz equation with point source:
%\begin{equation*}
    %\left( \nabla^2 + k_0^2 \right)\dyad{G}(\br , \br ') = - %\mathbb{I}_{3 \times 3} \delta(\br -\br '),
%\end{equation*}
%where $\mathbb{I}_{3 \times 3} = \mathrm{diag}(1,1,1)$ is the identity matrix of size $3$. 
%In two dimensions $\dyad{G}$ can be expressed via Hankel function of the first kind with zero order:
\begin{equation}
    \dyad{G}(\br , \br ') = \left(\mathbb{I}_{3 \times 3} + \frac{1}{k_0^2} \nabla \otimes \nabla \right) \frac{\mathrm{i}}{4} H_0^{(1)}\left( k_0 |\br -\br '| \right),
\end{equation}
where $\mathbb{I}_{3 \times 3} = \mathrm{diag}(1,1,1)$ is the $3 \times 3$ identity matrix.
The resulting Green's dyad is written in a compact form containing the in-plane as well as out of plane responses simultaneously. %Accordingly, we now consider the plane $y = 0$, so that $\br = (x, z)^{\mathrm{T}}$ and $\partial_y \equiv 0$.
We will assume that the plane of incidence is perpendicular to the rods, i.e $\br = (x, z)^{\mathrm{T}}$ [see Fig. \ref{fig1}(a)]. Multiplying \eqref{eq.ref2} by individual cylinder's polarizability $\dyadA$ and substituting $\br = \br_i$ yields dipole moment of the $i$-th scatterer:
\begin{equation}
    \bP (\br _i) = \dyadA \bE ^{\mathrm{inc}}(\br _i) + 4\pi k_0^2 \dyadA\sum\limits_{\substack{j \neq i}} \dyad{G}(\br _i, \br _j)\bP (\br _j),
    \label{eq.ref3}
\end{equation}
providing us with a system of $3N$ linear equations representing multiple scattering problem for an array of point sources.
The latter can be split into two independent systems associated with TM and TE polarizations; thus, only the values of $ P_y(\br _i)$ are needed for TE polarization that will be discussed further, which reduces the considered system to $N$ equations. In \eqref{eq.ref3}, the term $j = i$ is excluded to avoid the self-action. 
%The system of algebraic equations can be rewritten in the block matrix form as:
Resolving \eqref{eq.ref3} with respect to dipole moments, we obtain:
\begin{equation}
    \bP (\br _i) =\dyadtildeA_{i} \bE ^{\mathrm{inc}}(\br _i),
    \label{eq.ref4}
\end{equation}
where
\begin{equation}
    \dyadtildeA_{i}  = \sum\limits_{j=1}^N\left[ \mathbb{I}_{3\times 3} \delta_{i, j} - 4\pi k_0^2 \dyadA \dyad{G}(\br _i, \br _j) \right]^{-1} \dyadA e^{\mathrm{i}\mathrm{k}(\br_j-\br_i)}
    \label{eq.ref5}
\end{equation}
is the dressed polarizability of $i$-th cylinder, %$ \bP  = \left(0, P_y (\br _1), 0, \ldots , 0, P_y (\br _N),0 \right)^{\mathrm{T}}$, 
$ \bP(\br _i)~=~\left(0, P_y (\br _i), 0 \right)^{\mathrm{T}}$, 
and %$\bE ^{\mathrm{inc}} = E^0\left(0, e^{\mathrm{i}\mathrm{k}\br _1}, 0, \ldots , 0, e^{\mathrm{i}\mathrm{k}\br _N}, 0 \right)^{\mathrm{T}}$.
$\bE ^{\mathrm{inc}}(\br _i)~=~E^0\left(0, e^{\mathrm{i}\mathrm{k}\br _i}, 0 \right)^{\mathrm{T}}$.

\begin{figure}[t!]
\includegraphics[width=.8\columnwidth]{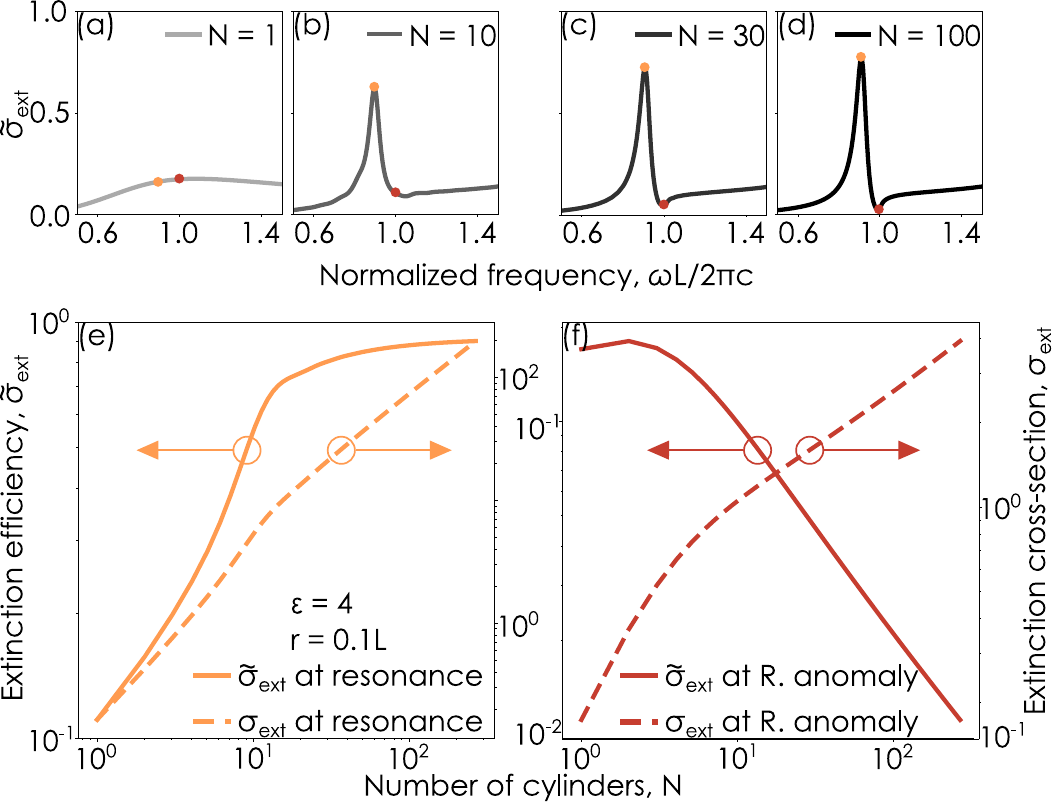}
\caption{(a)--(d) Extinction efficiencies for finite periodic array ($\eps = 4$, $r = 0.1 L$) with fixed number of rods $N = 1,~10,~30,~100$ respectively under normally incident TE-polarized illumination. Orange points correspond to the lattice resonance frequency and red points -- to the Rayleigh frequency for the infinitely long structure. (e), (f) Extinction efficiency (solid) / extinction cross-section (dashed) as functions of on $N$ at resonance and at the Rayleigh anomaly respectively.}
\label{fig2}
\end{figure}

The polarizability tensor $\dyadA$ of an individual circular cylinder %in the geometry presented on Figure \ref{fig1}(a) has only three diagonal components expressed as 
takes the form $\dyadA = \mathrm{diag} (\alpha_{xx},\alpha_{yy}, \alpha_{zz})$ with
$\alpha_{xx} = \alpha_{zz} = \mathrm{i}\frac{4}{k_0^2} a_1$, and $\alpha_{yy} = \mathrm{i}\frac{4}{k_0^2} b_0$.
The scattering coefficients $a_1$ and $b_0$ can be found in \cite[p.~301]{hulst1981light};
%In particular, the following expression for one of the coefficients was utilized during the computations:
for TE polarization analyzed in the following, only $b_0$ coefficient is relevant:
\begin{equation}
    b_0 = \frac{\sqrt{\varepsilon}J_0'(\sqrt{\varepsilon}k_0r)J_0(k_0r) - J_0(\sqrt{\varepsilon}k_0r)J_0'(k_0r)}{\sqrt{\varepsilon}J_0'(\sqrt{\varepsilon}k_0r)H^{(1)}_0(k_0r) - J_0(\sqrt{\varepsilon}k_0r)H_0^{(1)} {}' (k_0r)},
    \label{eq.ref6}
\end{equation}
where $J_0$ and $H_0^{(1)}$ are the Bessel and Hankel functions of the first kind and 0-th order respectively. 
% as functions of $k_0 r$ and $\sqrt{\varepsilon} k_0 r$. 

%To avoid increasing task dimensionality for TM polarization we will rewrite initial equation (\ref{eq.ref2}) for the total magnetic field parallel to the cylinder axis, which will similarly reduce the problem to a system of $N$ equations.
%The obtained values of the dipole moments at the scatterers locations can be substituted in following expression for the extinction cross-section 
The dipole moment values obtained by solving \eqref{eq.ref4} yield the extinction cross-section of the finite system:
\begin{equation}
    \sigma_{\mathrm{ext}} = \frac{4\pi k_0}{|\bE ^{0}|^2} \sum\limits_{i = 1}^{N} \Im \left[ \bP ^{*}(\br _i) \cdot \bE ^{\mathrm{inc}} (\br _i)\right].
    \label{eq.ref7}
\end{equation}
%Finally, we will examine average response from all nanoparticles -- the extinction efficiency of the cylinders array, i.e.,
Finally, we introduce the extinction efficiency defined as:
\begin{equation}
    \tilde{\sigma}_{\mathrm{ext}} = \frac{\sigma_{\mathrm{ext}}}{N L}.
    \label{eq.ref8}
\end{equation}

Figures \ref{fig2}(a)--(d) show a series of extinction efficiency spectra for finite arrays with different number $N$ of subwavelength dielectric cylinders illuminated with a normally incident TE-polarized plane wave.
While for a single cylinder, $N=1$, the extinction efficiency expectedly does not feature any resonant effects, sharp peaks and dips associated with the lattice resonances (orange dots) and Rayleigh anomalies (red dots) gradually appear for larger $N$.
The extinction efficiency evaluated at the lattice resonance frequency rapidly approaches the unitary value with growing $N$, Fig. \ref{fig2}(e), corresponding to quasi-perfect reflection of the incident field.
Conversely, at the first Rayleigh frequency, $\w L/ 2\pi c = 1$, the absolute extinction still grows with $N$, but the extinction efficiency vanishes in the limit of infinite array, $N \to \infty$. 
Thus, at the Rayleigh anomaly the system becomes more transparent in terms of extinction efficiency with increasing number of scattering elements.
More precisely, the extinction efficiency scales asymptotically as $\tilde{\sigma}_\mathrm{ext} \propto 1/\sqrt{N}$ [see~Fig.~\ref{fig2}(f)]. 
Such a slow convergence to the value predicted for an infinite system is caused by the lattice sum representing a conditionally convergent series of Hankel functions. 

\subsection{Effect of material loss}

The previous analysis ignores possible absorption in dielectric cylinders.
It is instructive now to examine how the presence of dissipation affects the results in light of the results on infinite arrays, which predict the transparency of the periodic system at the Rayleigh anomaly even in the presence of material loss \cite{DeAbajo2007}.
%In previous section material losses connected with permittivity imaginary part were excluded from consideration, which represented the equality of extinction and scattering for the system $\sigma_{\mathrm{ext}} = \sigma_{\mathrm{sct}}$. 
By introducing material loss we have to distinguish contributions from absorption and scattering to the extinction cross-section. 
%The absorption efficiency can be defined the similar way as extinction (\ref{eq.ref5}, \ref{eq.ref6}) when subtracting radiation losses, which is equal to replacement of incident field with total one
Absorption efficiency can be defined similarly to extinction, \eqref{eq.ref7} and \textup{{\normalfont(\ref{eq.ref8}}\normalfont)}, by replacing the background field with the total one:
\begin{equation}
    \tilde{\sigma}_{\mathrm{abs}} = \frac{4\pi k_0}{N L |\bE ^{0}|^2} \sum\limits_{i = 1}^{N} \Im \left[ \bP ^{*}(\br _i) \cdot \bE (\br _i)\right].
    \label{eq.ref9}
\end{equation}

\begin{figure}[t!]
\includegraphics[width=.8\columnwidth]{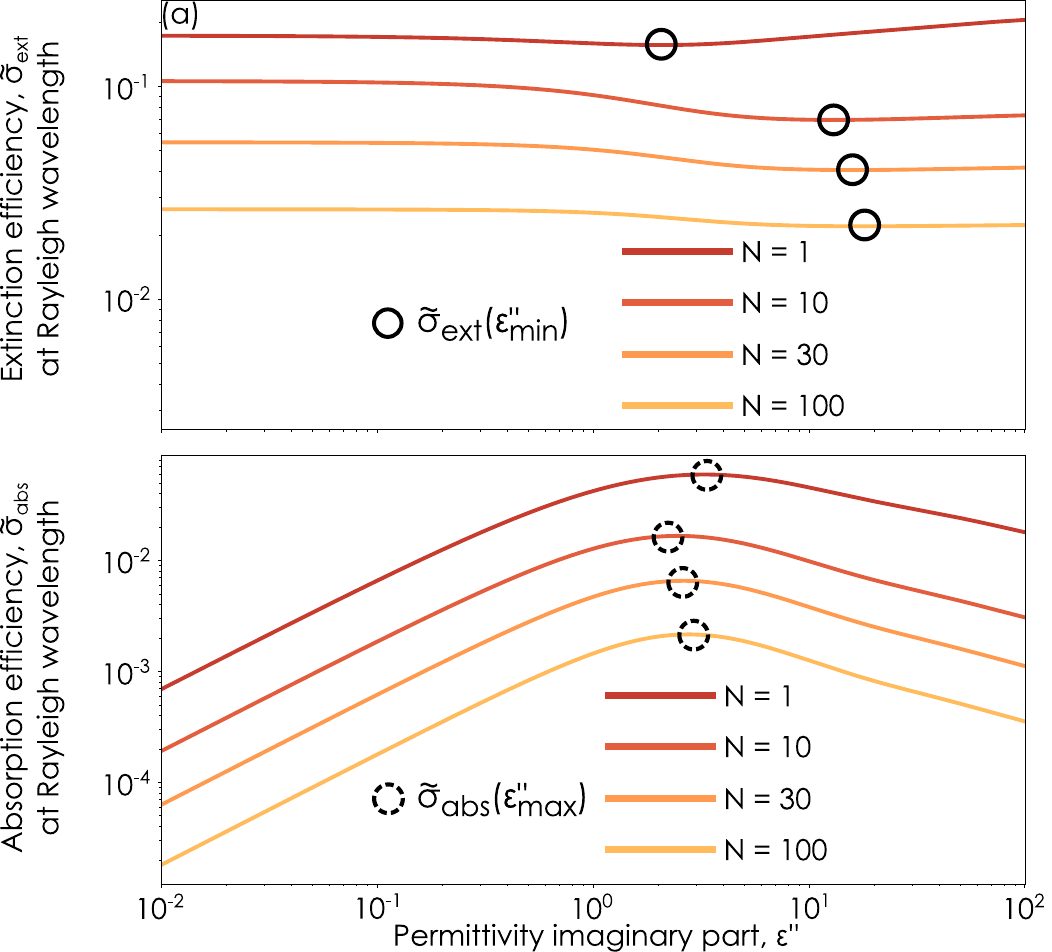}
\caption{(a) Extinction efficiency $\tilde{\sigma}_{\mathrm{ext}}$ and (b) absorption efficiency $\tilde{\sigma}_{\mathrm{abs}}$ for finite periodic array ($\mathrm{Re}\{\eps\} = 4$, $r = 0.1 L$) with fixed number of rods under TE-polarized illumination at angle $\theta = 0$ at the Rayleigh wavelength as a function of permittivity imaginary part $\Im [\eps] = \eps'' $. Solid circles show extinction minima and dashed circles -- absorption maxima. }
\label{fig3}
\end{figure}

%Figure \ref{fig3}(a) illustrates $\tilde{\sigma}_{\mathrm{ext}}$ behaviour for different imaginary parts of permittivity $\varepsilon = \varepsilon'  + \mathrm{i}\varepsilon'' $ at the Rayleigh wavelength for different sizes of an array. 
Figure \ref{fig3} presents the resulting extinction $\tilde{\sigma}_{\mathrm{ext}}$ and absorption $\tilde{\sigma}_{\mathrm{abs}}$ efficiencies as a function of the imaginary part of the cylinders permittivity $\varepsilon'' $ at the Rayleigh wavelength for a series of $N$. 
For each $N$ the absorption efficiency reaches a maximum at a certain value of $\varepsilon''$, Fig. \ref{fig3}(b).
At the same time, the increase of $\varepsilon'' $ suppresses the extinction efficiency below the initial value until it reaches a minimum, and then starts to increase approaching an asymptotic value, Fig. \ref{fig3}(a). 

The results in Fig. \ref{fig3}(a) are rather counter-intuitive.
An array of lossless cylinders becomes transparent at the Rayleigh anomaly in the limit of infinite array, $N \to \infty$ [see~Fig.~\ref{fig1}(b)].
A non-scattering optical system, such as an anapole nanodisk or an array of those, starts to scatter \emph{more} with increasing material dissipation in agreement with the optical theorem \cite{tretyakov2014maximizing, Fleury2014}. 
Nevertheless, increasing material absorption of the cylinders further suppresses extinction until the minimum value is reached.

It is worth mentioning that the absorptive performance of the system is entirely due to the properties of the individual particles, since the matrix denominator in \eqref{eq.ref5} contains factors consisting of two terms groups. While the second group describing array properties does not change with the material losses, the first one, represented through the inverse polarizability, determines the absorbing properties. 

%Interestingly, the extinction minimum shifts towards larger $\varepsilon'' $ as the size of the system continues to grow due to the lattice sum term in (\ref{eq.ref4}). 
%An infinite chain of lossless cylinders $N \to \infty$ at the Rayleigh wavelength is "transparent" \ref{fig1}(b), however increasing material absorption to the system even further will suppress extinction as minimum value of extinction will be reached in $\varepsilon''  \to \infty$.
%This is not what one intuitively expects from a non-scattering compact system. A scattering anapole system, for example, starts to scatter more with increasing internal dissipation due to the optical theorem. This  is what makes the results in Fig. \ref{fig3} rather counter-intuitive.

The inverse polarizability of a 2D scatterer takes the following form in quasi-static limit:
\begin{equation}
    \alpha_{yy}^{-1} = \xi' + \mathrm{i}\xi'' + \frac{\mathrm{i}}{4}k_0^2,
\end{equation}
where $\xi'$ is the electrostatic term, $\xi''$ represents absorption, and the last term describes the radiation correction \cite{tretyakov2014maximizing}. 
Utilizing the Taylor series expansion of \eqref{eq.ref6} at $k_0 r \to 0$ one can notice $\xi''$ vanishes in the limit of both small and very large material losses. 
As a result, in both limits the dominant contribution to the extinction efficiency is provided by scattering cross-section.
Indeed, $\sigma_{\mathrm{abs}} \propto \varepsilon''$ in the limit $\varepsilon'' \ll 1$, and $\sigma_{\mathrm{abs}} \propto 1/ \sqrt{\varepsilon''}$ in the limit $\varepsilon''  \gg 1$, according to results shown in Fig. \ref{fig3}(b).

%\rev{The system is kind of "transparent" at the anomaly with the normalized extinction approaching 0 in the limit $N \to \infty$. What is surprising here, is that adding material absorption to the system even further suppresses extinction towards 0.

\subsection{Numerical simulations: beyond point dipoles}

\begin{figure}[t!]
\includegraphics[width=.8\columnwidth]{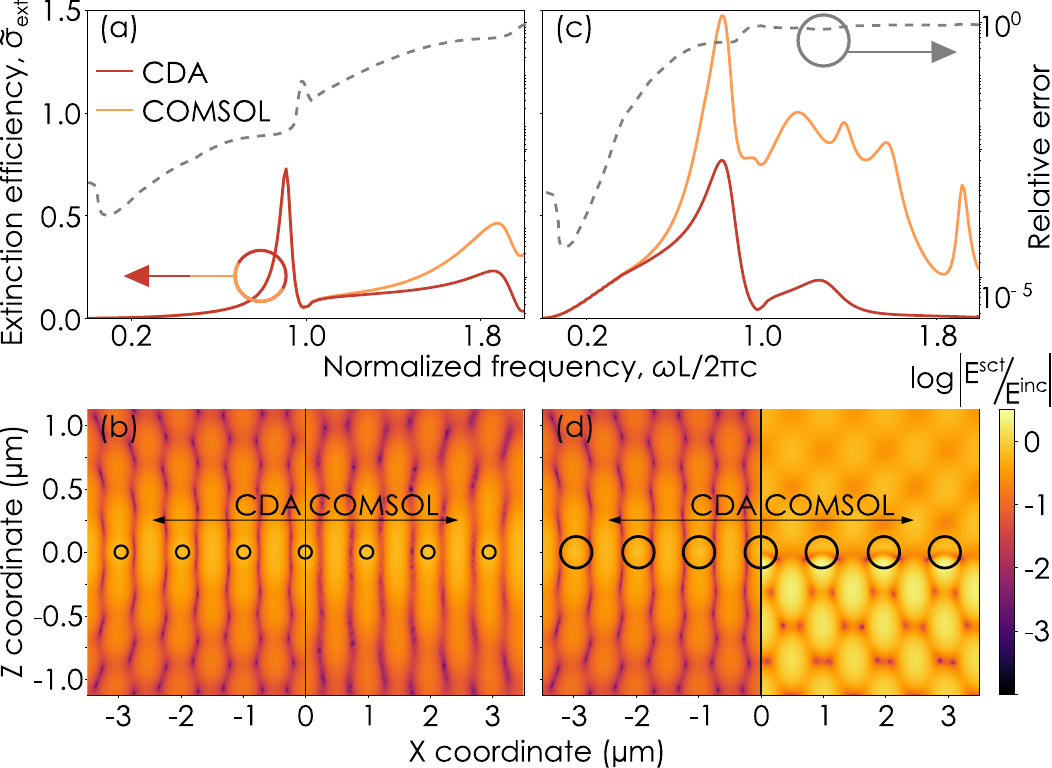}
\caption{(a) Comparison of the extinction efficiency $\tilde{\sigma}_{\mathrm{ext}}$ spectra obtained via CDA (red solid line) and numerical simulation performed in COMSOL Multyphysics (blue solid line) for $N = 30$ cylinders with $r = 0.1 L$. Dashed gray curve corresponds to relative error (log scale). (b) Scattered electric field absolute value as a function of the real space coordinates in the plane orthogonal to the cylinders axis for $N = 30$ given by CDA model and by COMSOL Multyphysics simulation at Rayleigh wavelength (black circles represent scatterers sizes and positions). (c) -- (d) Panels show analogous results for cylinders with $r = 0.25L$.}
\label{fig4}
\end{figure}

%\rev{To verify/justify the validity of coupled dipoles approximation used above, we perform a series of full-wave numerical simulations of ten finite array with ...}

To verify the validity of the coupled dipoles approximation used above, we perform a series of full-wave numerical simulations for $N = 30$ cylinders of different sizes in COMSOL Multiphysics. To that end, we calculate the scattering cross-section by integrating the energy flux over the boundary $\partial V$  between the Perfectly Matched Layers and cylinders as
\begin{equation}
    \sigma_{\mathrm{sct}} = \frac{1}{2I^{\mathrm{inc}}}\oint_{\partial V} \Re [\bE ^{\mathrm{sct} }\times \bH^{*,\mathrm{sct}}]\cdot \mathbf{n}\, dS,
    \label{eq.ref10}
\end{equation}
where $I^{\mathrm{inc}}$ is  energy flux
of the incident wave.
The absorption cross-section was obtained by the integration over scatterers  domain $V$ by the following expression:
\begin{equation}
    \sigma_{\mathrm{abs}} =\frac{1}{2I^{\mathrm{inc}}}\int_{V} \Re[\bj \cdot \bE^{*} ] d^3 \br ,
    \label{eq.ref11}
\end{equation}
where $\bj$ is total current density induced by total field $\bE $. 
Finally, extinction cross-section was determined as the sum of the absorption and scattering cross-sections, $\sigma_{\mathrm{ext}} = \sigma_{\mathrm{sct}} + \sigma_{\mathrm{abs}}$.

With the normalization given in \eqref{eq.ref8} taken into account, we obtained a coincidence of the results predicted by the numerical model and analytics with an accuracy of $10^{-2}$ for the both lattice resonance and Rayleigh wavelengths when the radii of the rod $r$ is ten times smaller than the structure's period $L$ according to Fig. \ref{fig4}(a). For the higher frequencies the results predicted by the models begin to diverge. 

The reason for the discrepancy in the results lies in the contributions of higher order multipolar resonances to the scattered field, which become significantly larger in the high frequency/low wavelength region or for the larger rod size and manifested in additional resonance peaks on Fig. \ref{fig4}(d) for $\w L / (2 \pi c) > 1$. The magnitude of each contribution is proportional to the corresponding scattering coefficient $a_l$ or $b_l$, $l = 0, 1, 2,\dots$, that is in turn a function of dimensionless size parameter $\sqrt{\varepsilon}k_0 r$. For the case $\sqrt{\varepsilon} k_0 r \ll 1$ coefficient $b_0$ remains dominant, immediately followed from Taylor series expansion of \eqref{eq.ref6}. 

Changes in the particle radii strongly affect the lattice resonances positions and quality factors, as they determined by the characteristic equation: %$\mathrm{det}~\Re \left\{\left(\dyadA^{\mathrm{dressed}}\right)^{-1}\right\} = 0$ 
\begin{equation*}
    \mathrm{det}~\Re \left\{\left(\dyadtildeA\right)^{-1}\right\} = 0,
    %\mathrm{det}~\Re \left\{1 / \dyadtildeA \right\} = 0,
\end{equation*}
which includes cylinders individual polarizabilities. However, the positions of Rayleigh anomalies are given by \eqref{eq.ref1} and do not depend on $r$ resulting in similarities for scattered field presented in Figures \ref{fig4}(b) and \ref{fig4}(c) predicted by the analytical model, slightly differing only in the magnitude. In addition, field profiles in Fig.~\ref{fig4}(b) are perfectly matched for both the CDA model and numerical simulation in COMSOL, which is violated by increasing the particle size leading to contribution of the higher multipoles and their interactions resulting in drastically different profiles in Fig.~\ref{fig4}(d).

\section*{Conclusion}
To conclude, we have studied theoretically and numerically the evolution of diffractive spectral features in finite arrays of subwavelength circular dielectric rods.
The results of the coupled dipoles approximation-based analytical model indicate that finite arrays do become transparent at the diffraction anomaly in terms of the normalized extinction. 
Furthermore, the presence of material absorption barely affects the transparency of finite arrays at the diffraction anomaly.
Finally, full-wave numerical simulations validate the results of the analytical model in the limit of subwavelength rods.
Our results could provide more insights into the diffractive effects in finite periodic systems.

\section*{Acknowledgments}
D.G.B. acknowledges support from Russian Science Foundation (grant No. 23-72-10005) and BASIS Foundation (grant No. 22-1-3-2-1).
The authors acknowledge the ITMO-MIPT-Skoltech Clover initiative.

\bibliography{Array}

\end{document}